# Time gated ion microscopy of light-atom interactions

P. Tzallas[1,2*], B. Bergues[3], D. Rompotis[4], N. Tsatrafyllis[1], S. Chatziathanassiou[1], A. Muschet[3,4], L. Veisz[3,5], H. Schröder[3] and D. Charalambidis[1,2*]

[1]*Foundation for Research and Technology—Hellas, Institute of Electronic Structure & Laser, PO Box 1527, GR71110 Heraklion (Crete), Greece.*

[2]*ELI-ALPS, ELI-Hu Kft., Dugonics ter 13, 6720 Szeged, Hungary.*

[3]*Max-Planck-Institut für Quantenoptik, D-85748 Garching, Germany.*

[4] *Deutsches Elektronen Synchrotron DESY, Notkestrasse 85, 22607 Hamburg, Germany*

[5] *Department of Physics, Umeå University, SE-90187 Umeå, Sweden.*

*Corresponding author e-mail address: ptzallas@iesl.forth.gr and chara@iesl.forth.gr

**Abstract.** The development of ultra-short intense laser sources in the visible and extreme ultraviolet (XUV) spectral range led to fascinating studies in laser-matter interactions and attosecond science. In the majority of these studies the system under investigation interacts with a focused laser beam, which ionizes the system. The ionization products are usually measured by devices, which spatiotemporally integrate the ionization signal originating from the entire focal area, discarding in this way valuable information about the ionization dynamics that take place in the interaction volume. Here, we review a recently developed approach in measuring the spatially resolved photoionization yields resulting from the interaction of infrared (IR)/XUV ultra-short laser pulses in gas phase media. We show how this approach enables a) the in-situ focus diagnostic, b) quantitative studies of linear and non-linear ionization processes in the IR/XUV regime, c) single-shot XUV-pump-XUV-probe studies and d) single-shot 2nd-order XUV autocorrelation measurements. The article has been published in *J. Opt.* **20**, 024018 (2018) (doi.org/10.1088/2040-8986/aaa326).

## 1. Introduction

In its half century long history, the investigation of strong-field and multi-photon laser matter interactions went through several refinements. Improvements are based both on the advances in the driving laser technology as well as in the development of specialized measurement instrumentation and methods. Examples of the first, include the shortening of the laser pulses to few cycles [1-6], CEP control [7], high pulse contrast and increased repetition rates at high pulse energies. Imaging techniques such as Velocity Map Imaging (VMI) [8,9,10,11] and slice imaging [12,13], flat field spectrometers, and coincidence techniques as Reaction Microscopy [14], are representatives of advanced instrumentation that allowed for detailed studies and deeper insight in the phenomena under investigation.

Two parameters intrinsic to the laser sources used in strong field and multi-photon ionization experiments have prevented for longer time the observation of several phenomena and/or the details of the effects under investigation. Those are the temporal and spatial distributions of the laser (or other source) electric field amplitude. In contrast to linear processes, a challenge in the study of non-linear interactions is the detection of an event at well-defined time and space. The measuring devices of the interaction products are commonly averaging both in time and space, while most of the phenomena studied depend on radiation parameters varying in time and space. Thus, the spatiotemporal laser field distribution in combination with the averaging detection devices prevents in many cases the study of subtle intricacies of light-matter interactions.

The effect of the temporal variation of the laser field envelope is more pronounced for the long pulses [15]. The observation of the interaction products is averaged over many cycles and there is no way to limit it to few cycles around the pulse peak intensity of the driving field. This is because the atoms/molecules are fully ionized at the rising edge of the multi-cycle laser field. This has prevented for long time the observation of pure non-perturbative effects such as high-order harmonic generation [16,17,18], above threshold ionization [18,19,20,21,22,23,24,25], and direct double ionization [18, 26,27,28,29]. The reduction of the pulse duration reduces the "averaging" time and has allowed the observation of the above effects at the resulting limited interaction time interval. In laser interactions with condensed matter, the observation of such effects was considered not possible as the sample would be destroyed at the required intensities. The development of single-cycle pulses [30] has recently allowed the observation of such effects in crystals [31] thanks to the ultra-short interaction time that has become available. Similarly the observation of direct two photon double ionization of atoms became possible only recently thanks to the generation of very short XUV pulses [32]. Few other effects such as relativistic interactions of outer shell electrons with laser fields or ionization stabilization [18,33] are still not observable due to the temporal evolution of laser pulses.

Spatial integration of the interaction products has been a further obstacle for the detailed study of strong field and multi-photon processes. "Volume effects" due to averaging of ionization products over the spatial intensity distribution in the focal volume has obscured the direct observation of ionization saturation and depletion of the atomic/molecular ground state, preventing straightforward quantitative measurements and reducing the contrast of measurements [34,35]. Successful attempts for spatially resolved multi-photon ionization have been demonstrated already in the 90's [36, 37,38, 39]. Recently an advanced approach for spatially resolved measurements has been developed and implemented in the IR [40] and XUV spectral range [41,43,44]. It is based on the "Ion Microscope" (IM), which is an ion imaging device facilitating the observation of the spatial distribution of the ionization products produced in a focused beam as a function of their mass over charge ratio $m/q$, i.e. it is an ion mass spectrometer with spatial imaging capability. Since the ionization process occurs in a focused radiation beam, an immediate advantage in using this device is the observation of local intensity effects within the laser focus, saturation and aberrations being two of those. It therefore facilitates quantitative measurements, which are complicated when volume averaged signals are analyzed. It substitutes imaging approaches that detect photons produced by non-linear processes and used in pulse metrology, which is beneficial in particular in the XUV spectral region. A notable application of the device is in the investigation of the intrinsic spatiotemporal properties of the radiation source. In the majority of the studies, the ionization occurs and is observed at the focus of the beam of a radiation source. Thus, the ionization area is an image of the radiation source itself. The recorded images of the spatial ion distribution at the focus of the radiation, reflect the spatial distribution of the radiation source, this being an HHG or FEL or a laser source and thus detailed information about the processes occurring at the source can be extracted.

In this work we review recent results obtained using the ion microscopy technique in different laboratories, i.e. FORTH, MPQ and FLASH using three different types of sources, namely lasers, HHG and FEL sources. A description of the operational principle (section 2) of the device is followed by examples of its use in linear XUV interactions, non-linear IR and XUV interaction (sections 3, 4), while a separate chapter is dedicated to applications of the device in XUV-pump-XUV-probe measurements and XUV pulse metrology (sections 5, 6).

## 2. Operation principle of the Ion Microscope

A schematic of the IM is shown in Fig. 1a. The spherical mirror (SM) shown in Fig. 1a is focusing the back reflected beam into the interaction region. This arrangement is very useful in experiments performed with few-cycle laser pulses and pulses in the XUV spectral range. The use of a lens is also convenient for focusing multi-cycle UV to mid-IR laser pulses. The interaction region is filled with the gas under investigation by means of a needle valve, which provides a continuous gas flow or by using a piezo pulsed nozzle. The ion distribution produced at the focus of the laser beam (located in the object plane) is mapped onto a position-sensitive detector located at the image plane. As opposed to a velocity mapping spectrometer that images the momentum distribution of the charged particles, the IM reveals their spatial distribution in a magnified high-resolution manner. The ions generated at the focus of the laser beam are first accelerated by an electric field applied between the repeller (being floated at voltages $V_{rep}$) and extractor electrodes (being floated at voltages $V_{ext}$) placed at a distance ≈ 0.5 cm from the repeller. A first electrostatic lens (EL1 placed at a distance ≈ 2 cm from the extractor) images the spatial extent of the ion cloud on an intermediate ion image plane with a small magnification factor $M_1$ ranging from 5 to 7. This intermediate image is located at the focal plane of a second electrostatic lens (EL2 placed at a distance ≈ 16 cm from the repeller) that projects a further magnified (a magnification factor of $M_2$ ranging from 20 to 30) image onto the detector consisting of a pair of MCPs (placed at a distance ≈ 50 cm from the repeller) and a phosphor screen. The image of the ion cloud that appears on the phosphor screen is magnified by a factor $M = M_1 \times M_2$ and is recorded with a CCD camera. We note that as the images are the projections of the three dimensional ion distributions with cylindrical symmetry in the observation plane, an Abel transformation is required for obtaining the actual size of the produced ion distribution. Since fragments with different charge to mass ratios have different flight times, gating the detector with a few-tens of *ns* time window enables mass selection of individual charge states. Time-of-flight (TOF) mass spectra showing the arrival times on the MCP of the ions generated by the interaction of the laser beam with the atoms and molecules are shown in Fig.1b and 1c, respectively. The shaded green area (G) in the mass spectra shows a temporally controlled gate which is applied on the MCP for the selection of the TOF mass peak ($m / q$) the ion distribution of which is going to be imaged. Fig. 1d shows the single-shot ion distribution recorded at the focus of an XUV beam which contains the up-converted harmonic frequencies (11th to 15th) of a

≈ 30 fs long infrared laser pulse. The XUV beam was focused into the interaction region (filled with Argon gas) with a spherical gold mirror (SM) of 5 cm focal length. Since the distribution is produced through a 1-XUV-photon ionization process of Argon, it corresponds to the intensity distribution of the XUV radiation in the focal area. In this measurement the XUV focal spot diameter is found to be ≈ 2 μm, the spatial resolution of the device being ≈ 1μm.

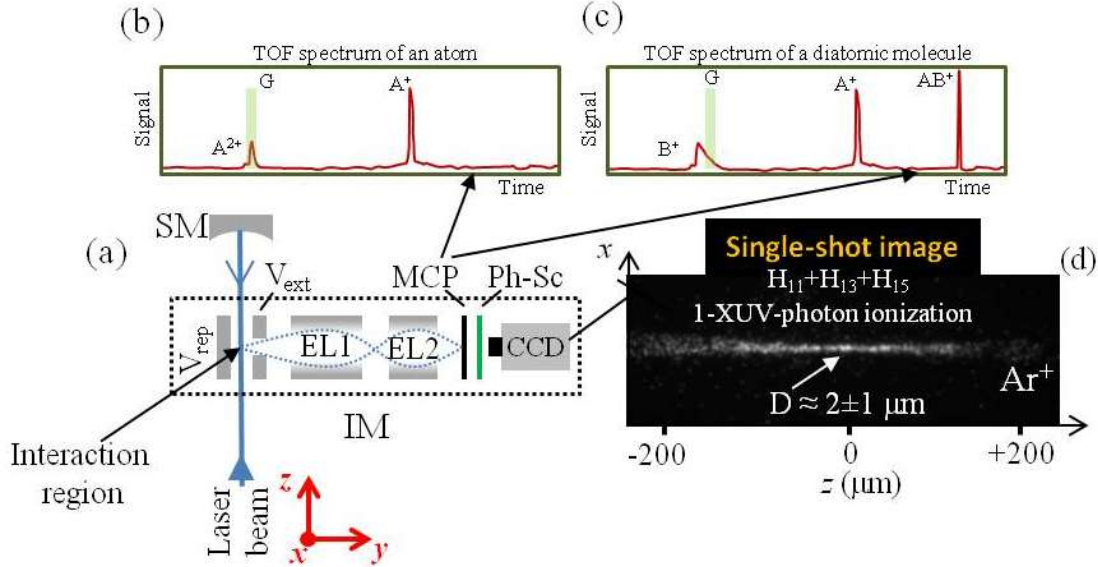

**Figure 1.** (a) A schematic of the operation principle of the IM. The IM consists of the interaction region where a laser pulse is focused and the electrostatic lenses EL1, EL2 used for magnifying and imaging the ion distribution produced in the interaction region on the MCP detector coupled with a phosphor screen (Ph-Sc). The image is recorded by a CCD camera. b) and c) A sketch of TOF mass spectra produced by the interaction of the focused beam with atoms (A) and diatomic molecules (AB), respectively. The shaded green area (G) in the mass spectra shows a temporally controlled gate which is applied on the MCP for imaging the ion distribution of a specific mass peak (m/q). d) Single-shot ion distribution recorded at the focus of an XUV beam which contains the harmonics from 11th to 15th. The ion distribution was produced through 1-XUV-photon ionization of Argon.

The spatial resolution of the IM strongly depends on the density of ions (space charge effects) and the DC electric field strength applied in the TOF acceleration region ($V_{rep} - V_{ext}$). Optimum spatial resolution values can be reached by minimizing the density of ions and maximizing the DC electric field strengths in the TOF acceleration region. Ion trajectory calculations using SIMION for a particular IM configuration [42] showed that for operation voltages $V_{ext}$ ≈ 10 kV, $V_{rep}$ ≈ 8.5 kV, $V_{EL1}$ ≈ 6.3 kV and $V_{EL2}$ ≈ 8.5 kV, the magnification can be in the range of M ≈ 150 with a spatial resolution in the range of ≈ 270 nm. However, mainly due to space charge effects and laser pointing instabilities, the best reported spatial resolution was in the range of ≈ 1 μm [43,44]. In case of multiple ion production and according to the results and the experimental conditions used in ref. 44, an ion density ($\rho_{ion}$) which maintains a spatial resolution $< 1$ µm is $\rho_{ion} \sim 10^{11}$ ions/cm$^3$ per shot. This value can be farther increased to the level of $\rho_{ion} \sim 10^{14}$ ions/cm$^3$ per shot by increasing the operation voltages of the IM. Also, ion distribution images free of space charge effects can be recorded by measuring single ions in a multi-shot experiment using high repetition rate laser systems.

## 3. Spatially resolved ion distribution produced in the non-linear IR regime

As discussed above, the ion microscope is an ion time of flight spectrometer that offers the possibility to measure the spatial ion distribution of different charge states. This allows for detailed studies of the intricacies of multi-photon and/or strong field ionization. Such distributions measured for the ionization of Argon atoms by focused 800nm, 25 fs laser pulses are shown in Fig. 2 for three different laser pulse energies. The focal areas are clearly observable for the charge states 1+, 2+ and 3+.

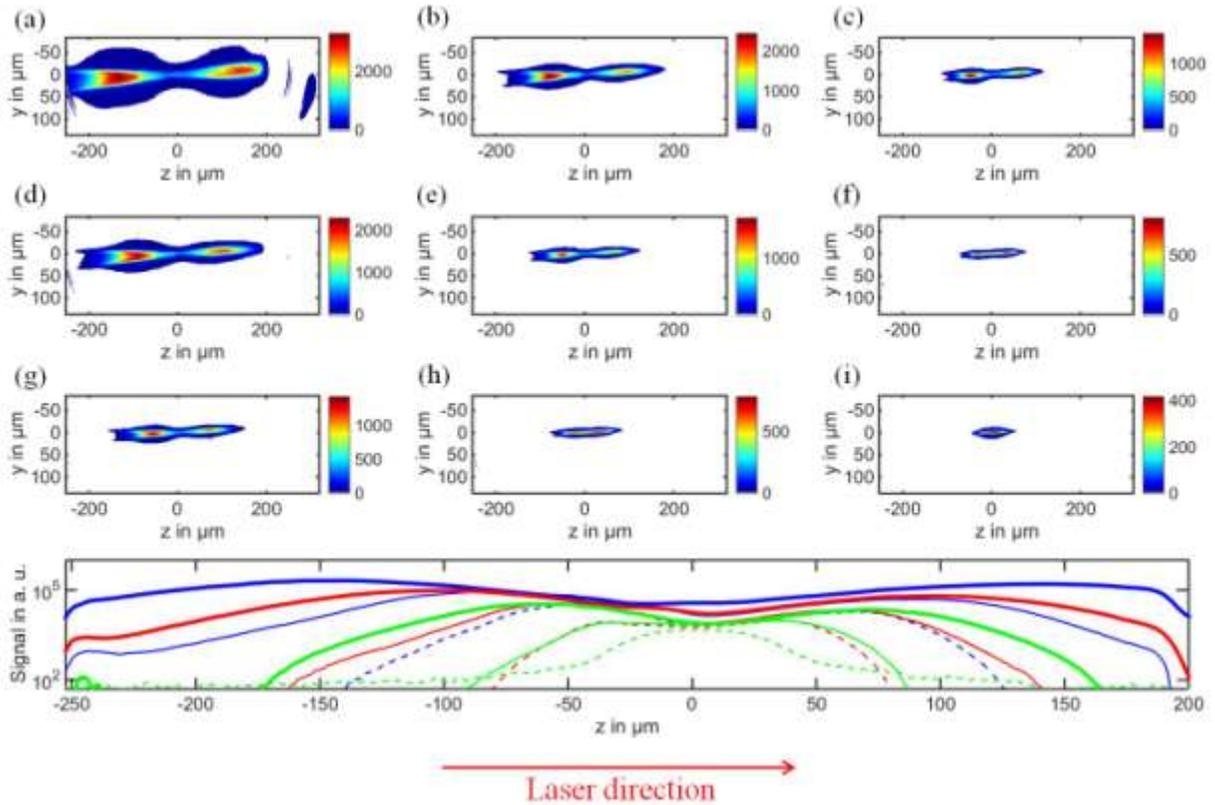

**Figure 2.** False color raw images of the charge states Ar$^{1+}$ (a, d, g), Ar$^{2+}$ (b, e, h) and Ar$^{3+}$ (c, f, i) produced in the focus of a Ti:Sapphire laser beam with a pulse energy of 247 µJ (a, b, c), 109 µJ (d, e, f) and 47 µJ (g, h, i). The projection in x and y direction of the same data are shown in (j), where the colors blue, red and green show the projections of 247 µJ, 109 µJ and 47 µJ respectively. The line types thick solid, thin solid and dashed denote the ionization states Ar$^{1+}$, Ar$^{2+}$ and Ar$^{3+}$ respectively.

The first striking observation is that in all distributions, except the $Ar^{3+}$ distribution produced by a 47 µJ pulse energy, the relative ion number is reduced at the position where the laser intensity is highest (at the focal plane). This is due to saturation of the ionization process i.e. depletion of the parent charge states producing these ions. At 247 µJ and 109 µJ the ionization of Ar, $Ar^{1+}$ and $Ar^{2+}$ is saturated. At 47µJ the ionization of Ar and $Ar^{1+}$ is saturated, while the ionization of $Ar^{2+}$ is not. The 1D distributions in Fig. 2 are projections of the 2D distributions on the propagation axis (z-axis), i.e. integrated signal over the y-axis. In spatially not resolved measurements it is well known that, in the volume integrated ion signal, saturation only manifests itself by a change in the slope of the log of the ion yield vs the log of laser intensity curves. For Gaussian beams the slope reaches a value of 3/2 beyond the saturation intensity. Thus, for few photon ionization the determination of the saturation intensities becomes ambiguous. Low spatial resolution non-imaging methods have been applied in the past [36 and references therein]. The ion microscope represents a significant advancement for spatially resolved measurements with higher spatial resolution, increasing the accuracy of quantitative studies of multiphoton or strong field processes.

A second observation in Fig. 2 is the asymmetry of the observed 2D images with respect to the focal plane, which arises from aberrations of the focusing system. This demonstrates that the IM is a valuable tool for evaluating the quality and properties of the focused beam. Figure 3 depicts a small portion (relative to the Rayleigh range) of focal area around the beam waist, imaged using four argon charge states. The reduction of the size of the beam waist with charge state is obvious and originates from the higher order of nonlinearity of the ionization process leading to the higher charge states.

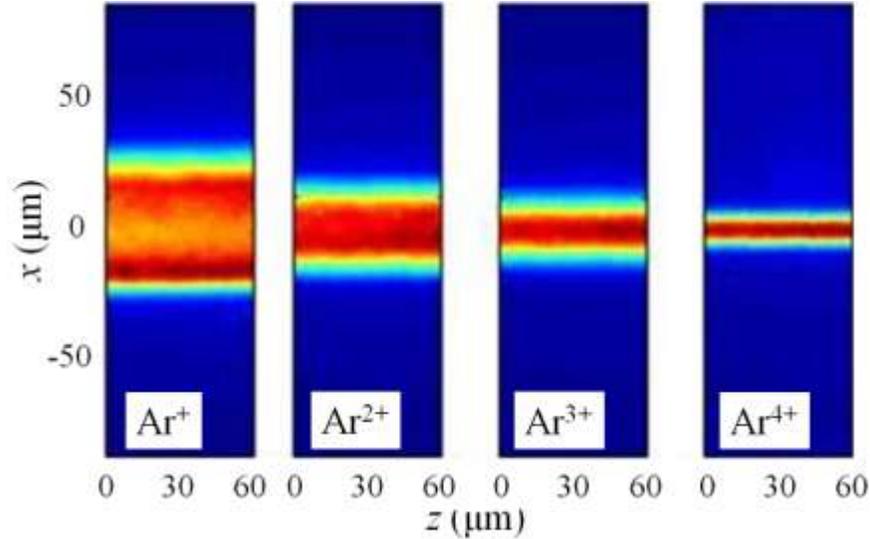

**Figure 3.** False color raw images of the charge states $Ar^{1+}$ to $Ar^{4+}$ as recorded with the IM at the beam waist. The images were recorded over 60000 shots. From ref. [40]

**4. Spatially resolved ion distribution produced in the linear and non-linear XUV regime**

Radiation in the XUV spectral range is currently generated by laser driven XUV [45] and FEL sources [46]. In both cases the measurement of the XUV intensity distribution at the focus of the XUV source is of particular importance as it can provide a natural XUV intensity scale, quantitative information about the properties of the XUV source and detailed information about the ionization dynamics of the system under investigation. In the following sections (4.1 and 4.2) we will review recent experiments performed in these directions using laser driven and FEL XUV sources.

**4.1. Gas phase laser driven XUV sources**

The results presented in this section were obtained using XUV radiation produced by the interaction of Xenon gas with a 10 Hz repetition rate high power Ti:Sapphire $\tau_L \approx 33$ fs laser pulse focused in the Xenon gas using a 3 m long focal length lens. The XUV beam was spectrally filtered and the comb of harmonics 11th-15th were selected for this study [44]. The harmonics were focused into the target gas jet (filled with Ar or Helium) with a spherical gold mirror (SM) of 5 cm focal length. The energy of the XUV radiation in the interaction region of the IM was obtained from the measured pulse energy using an XUV calibrated photodiode taking into account the reflectivity of the gold spherical mirror (SM shown in Fig. 1).

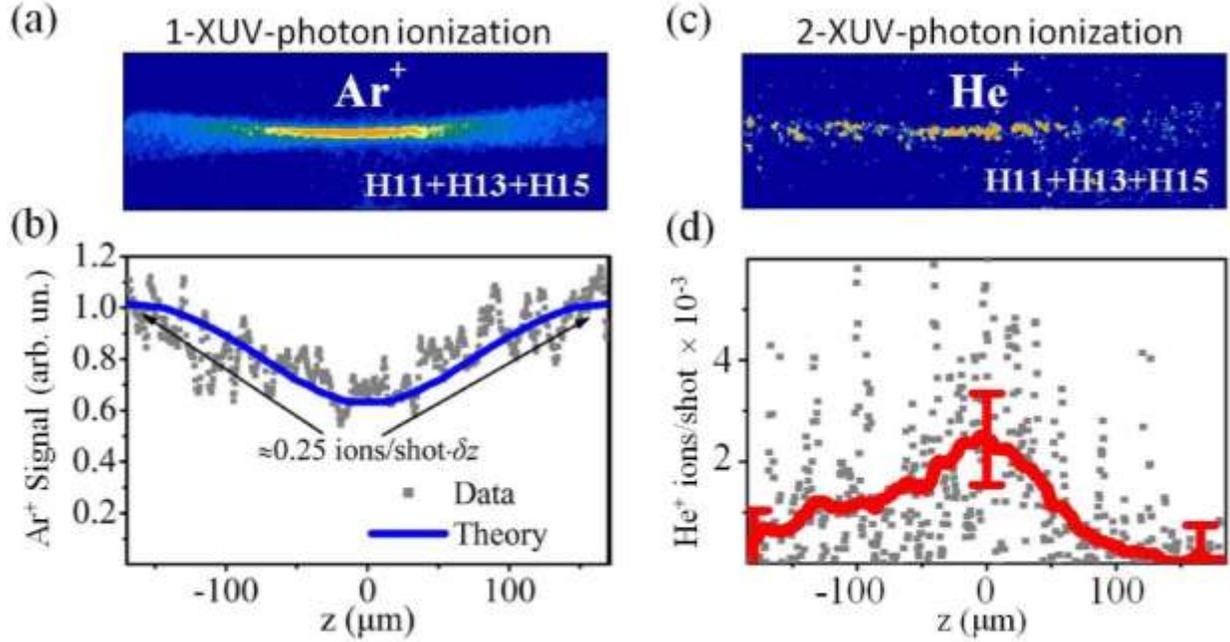

**Figure 4.** (a) Ar$^+$ spatial distribution at the focus of the XUV beam produced by a 1-XUV-photon ionization process. Images of 600 shots were accumulated. The images were recorded using XUV pulses of energy 20 nJ/pulse. (b) Volume integrated Ar$^+$ signal at each position of the XUV focus (gray points). The blue line shows the theoretically calculated signal. (c) He$^+$ spatial ion distribution at the focus of the XUV beam produced by a 2-XUV-photon ionization process. Images of 15000 shots were accumulated. The images were recorded using XUV pulses of energy 90 nJ/pulse. (d) Number of generated He$^+$ ions per shot in the XUV focal area. The red line is the 170-point running average of the raw data (gray points) and used only to show the dependence of the mean value of He ions on *z*. The error bars represent one standard deviation of the mean value. Fig. from ref. [44]

Also, the number of ions was obtained by calibrating the detector for single ion detection [44]. The spatial ion distribution at the focus of the high-harmonic beam was recorded using an IM with ≈ 1 μm spatial resolution. Spatially resolved 1-XUV- and 2-XUV-photon ionization products of Argon (Fig. 4a) and Helium (Fig. 4c) were observed. The Ar$^+$ and He$^+$ ion distributions were recorded using XUV radiation with energy ≈ 20 nJ/pulse and ≈ 90 nJ/pulse, respectively, while all other experimental conditions were the same. In both cases the duration of the XUV pulse is considered to be $\tau_{XUV} = \tau_L/\sqrt{n} \approx 15$ fs (when $n$ = 3-5 is the effective number of non-linearity of the harmonic generation process for harmonics close on the plateau region [47]).

### 4.1.1. Spatial ion distribution produced by 1-XUV-photon ionization

The spatial ion distribution produced by a 1-XUV-photon ionization process at the focus of the XUV beam can be used a) for the characterization of the XUV focus and b) for obtaining the XUV pulse energy ($E_{XUV}$), the 1-XUV-photon ionization cross section ($\sigma^{(1)}$) and the atomic density ($\rho$) in the interaction region.

- *XUV focus characterization*: Since the 1-XUV-photon signal is proportional to the intensity of the XUV pulse, the image of the ion distribution provides detailed information about the XUV intensity distribution and the XUV beam radius ($R(z)_{XUV}$) in the focus. The latter, in combination with independent measurements of the XUV energy ($E_{XUV}$) and $\tau_{XUV}$, can give (using the equation $I_{XUV}(z) = E_{XUV}/(\tau_{XUV} A(z))$ where $A(z) = \pi R(z)^2_{XUV}$) an accurate value of the intensity of the XUV pulse along the focus. We note that due to the limited spatial resolution of the IM the accuracy in obtaining $I_{XUV}(z)$ is reduced as $z \to 0$, where the spatial resolution is comparable with the size of the focal spot diameter.

- *Measurement of the $E_{XUV}$, $\sigma^{(1)}$ and $\rho$*: The equation which connects the quantities $E_{XUV}$, $\sigma^{(1)}$ and $\rho$ in a single photon ionization process is

$$N^{ions} = \sigma^{(1)} \cdot \delta z \cdot \rho \cdot E_{XUV}/(\hbar \omega_{XUV}) \tag{1}$$

where $N^{ions}$ is the number of ions generated in a segment of length $\delta z$, $\rho = \rho_{Ar} = n_{Ar}/V$ ($n_{Ar}$ is the number of atoms in the interaction volume $V$) is the atomic density and $\omega_{XUV}$ is the carrier frequency of the XUV radiation. From the above equation the quantities of $N^{ions}$ and $\delta z$ can be obtained from the image of the ion distribution. It is evident that $\rho$ is the only quantity which remains to be experimentally determined in order to have a direct correspondence between $E_{XUV}$ and $\sigma^{(1)}$. A way to bypass determination of $\rho$ is to work in the ionization saturation regime of the single photon ionization. In this case

$$N^{ions} \propto [1 - \exp(\frac{I_{XUV}(z)}{I^{sat}_{XUV}})] V(z) = [1 - \exp(\frac{E_{XUV} \sigma^{(1)}}{(\hbar \omega) A(z)})] V(z) \tag{2}$$

where $V(z) = \delta z \cdot A(z)$ and $I^{sat}_{XUV} = (\hbar \omega_{XUV})/(\tau_{XUV} \sigma^{(1)})$ is the ionization saturation intensity [36,48,49,50]. The ionization saturation regime appears in the ion distribution as a reduction (associated with the term $\exp(\frac{E_{XUV} \sigma^{(1)}}{(\hbar \omega) A(z)})$) of ion yield for intensities $I_{XUV}(z) > I^{sat}_{XUV}$ (Fig. 4b). The above relation, which establishes a direct connection between $E_{XUV}$ and $\sigma^{(1)}$, was used for the theoretical calculation shown with the blue line in the lower panel of Fig. 4a which is found to be in excellent agreement with the experimental data. In this way, using the values $N^{ions}$, $\hbar \omega_{XUV}$, $A(z)$ and $V(z)$ measured by the ion distribution image and the value of $E_{XUV}$ measured by a calibrated XUV photodiode we can obtain the value of $\sigma^{(1)}$ which is found to be $\approx$ 33 Mb and vice versa using the value of $\sigma^{(1)}$ [51,52,53] we can obtain the value of $E_{XUV}$ which is found to be

≈ 20 nJ/pulse (details can be found in ref. [44]). We note that in the region of $z \approx \pm 160$ μm, where the signal is close to unity, the XUV intensity is well below the ionization saturation. In this regime $\rho$ can be obtained using eq. 1 and the measured values of $E_{XUV}$ and $\sigma^{(1)}$.

Summarizing, the quantities that can be obtained from an ion distribution image produced at the XUV focus by a 1-XUV-photon ionization process are the: I) $I_{XUV}(z)$ (for a known $\tau_{XUV}$), II) $E_{XUV}$ (for a known $\sigma^{(1)}$), III) $\sigma^{(1)}$ (for a known $E_{XUV}$) and IV) $\rho$ (for a known $E_{XUV}$ or $\sigma^{(1)}$), subject to which of them are already known from previous or other measurements.

**4.1.2. Spatial ion distribution produced by 2-XUV-photon ionization**

The spatial ion distribution produced by a 2-XUV-photon ionization process at the focus of the XUV beam can be used a) for the measurement of the order of the non-linearity of the ionization process and b) for obtaining the value of the 2-XUV-photon ionization cross section ($\sigma^{(2)}$).

- *Measurement of the order of non-linearity in the XUV region:* A conventional way for measuring the order of non-linearity of an ionization process relies on the dependence of the ion yield on the energy (and consequently the intensity) of the driving laser field. These measurements require the acquisition of large number of shots at different values of the laser intensity. In laser driven gas phase XUV sources the change of the XUV energy occurs by changing the gas pressure in the harmonic generation medium. Although this procedure can give accurate results when the duration XUV pulses are in the *fs* range, when we are dealing with pulses in the attosecond regime the procedure can lead to safe results only for small pressure changes, which do not influence the spectral amplitude/phase modulation of the generated XUV radiation and hence the duration of the attosecond pulses.

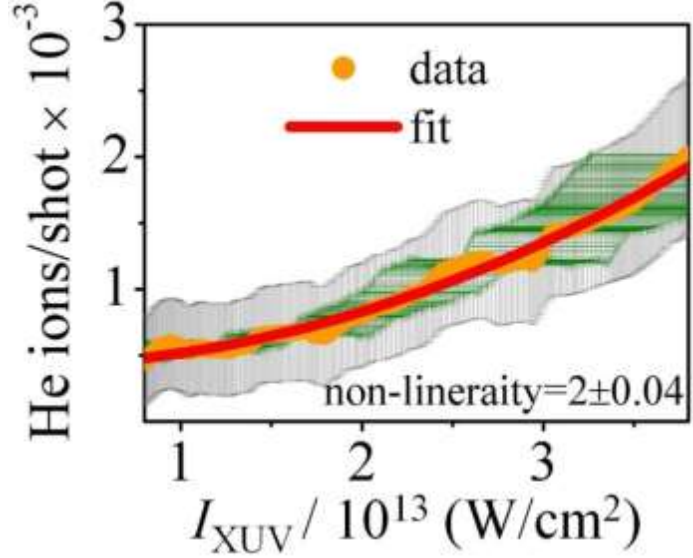

**Figure 5.** Dependence of the He$^+$ ion number per shot per $\delta z$ on the intensity of the XUV radiation (orange dots). The dependence was obtained using the values of the red-line of Fig. 4d. The red line shows the non-linear fit of the data. The gray and green error bars represent one standard deviation of the mean value of He$^+$ and $I_{XUV}$, respectively. From ref. [44].

A spatially resolved ion distribution does not suffer from such complications because it is recorded without any change in the experimental conditions. The He$^+$ yield shown in Fig. 4d (resulting from the He$^+$ image shown in Fig. 4c, produced through 2-XUV-photon of He) in combination with the measured $I_{XUV}(z)$ confirms a quadratic dependence of the He$^+$ yield on $I_{XUV}$ (Fig. 5) as is expected from the Lowest Order Perturbation Theory (LOPT) (for details see ref. [44]).

- *Measurement of the $\sigma^{(2)}$ in the XUV region*: As it is noted at the end of the section 3.1.1. the cross sections of the multi-photon ionization processes were measured in the IR and UV spectral regions using the most of the times approaches (similar to those described in section 3.1.1), which rely on the saturation of the multi-photon ionization process [36,48,49,50]. These approaches are hardly applicable in the XUV spectral region due to the limited number of sources, which can lead to saturation of the multi-photon ionization process. To our knowledge, there is only one measurement of $\sigma^{(2)}$ of He in the 20 eV photon energy range and this has been achieved using a free electron laser (FEL) source [54].

Here, the $\sigma^{(2)}$ of He in the 20 eV photon energy range was obtained using the data of Fig. 5 and from the well-known LOPT equation $\sigma^{(2)} = (\hbar\omega_{XUV})^2 N^{ions}/(I_{XUV}^2 \cdot \tau_{XUV} \cdot \rho \cdot V(z))$. In this way $\sigma^{(2)}$ can be obtained over a large XUV intensity range. In the XUV intensity range from $7 \times 10^{12}$ W/cm$^2$ to $4 \times 10^{13}$ W/cm$^2$ it has been found that $\sigma^{(2)} \approx 5 \times 10^{-52}$ cm$^4$s [44], which is in reasonable agreement with the theoretical [55,56,57] and experimental values obtained using FEL sources [54].

Summarizing, the ion distribution image produced at the XUV focus by a 2-XUV-photon ionization process can provide quantitative information about: I) The order of non-linearity of the process (for known $I_{XUV}(z)$) and II) $\sigma^{(2)}$ (for a known $I_{XUV}(z)$ and $\rho$). In both cases, as described in the previous section, $I_{XUV}(z)$ and $\rho$ can be experimentally obtained by the ion distribution image produced by a 1-XUV-photon ionization process. Nevertheless, we note that although the experimental conditions used for obtaining the Ar$^+$ and He$^+$ ion distributions were identical, the change of the gas in the IM interaction region introduces an error on the obtained results. This error is associated with the density of He and Ar atoms in the IM interaction region which was considered to be equal i.e. $\rho_{Ar} = \rho_{He}$ (for details see ref. [44]). A way to eliminate this error is to perform studies in a single gas medium using multi-photon multiple ionization process. For example in the 20 eV photon energy range the $\sigma^{(2)}$ of the two-XUV-direct double ionization Ar can be obtained using the Ar$^{2+}$ and Ar$^+$ spatially resolved distributions.

### 4.1.3 Mapping the high-order harmonic generation process

The spatially resolved image of the ion distribution produced by 1-XUV-photon process apart from providing detailed information about the XUV intensity distribution at the focus, it can also be used to diagnose the properties of the XUV source. For example, in the framework of the recollision picture [45,58,59,60,61,62 and references there in] it is well known that two electron trajectories (the Short (S) and the Long (L)) contribute to the emission of the plateau high-order harmonics in the laser driven gas phase media. The long-trajectory harmonics have different divergence and phase than the short-trajectory ones. The long-trajectory harmonics depict an annular shape beam profile with larger divergence than the short-trajectory harmonics which appear to be confined at the center of the XUV beam [63]. According to the relation $\varphi_q^{L,S} \propto \tau_q^{L,S} U_p$ (where $U_p$ is the ponderomotive energy of the electron) the phase of each harmonic order $q$ is proportional to the intensity of the driving field $I_L$ (additional information can be found in ref. [62,45] and references there in). The long- and short-trajectory beam profile can be measured in the far field of the XUV beam, using an XUV beam profiler, the knife edge technique or the spatially resolved ion distribution produced 1-XUV-photon ionization process (Figs. 6a,b). Since the phase difference between the long- and the short-trajectory harmonics ($\Delta\varphi_q^{L,S} = \varphi_q^S - \varphi_q^L$) depends on $I_L$ a single- and double-peak structure is expected to appear along the z-axis of the ion distribution when $\Delta\varphi_q^{L,S} = 2n\pi$ and $\Delta\varphi_q^{L,S} = (2n+1)\pi$ (where n=0,1,2...), respectively.

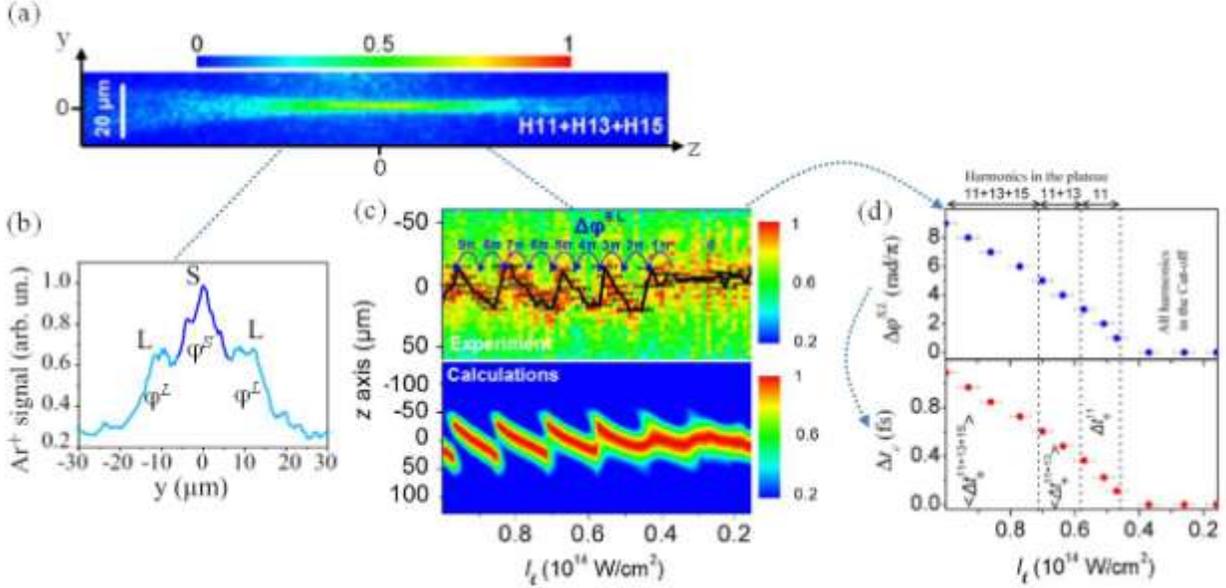

**Figure 6.** (a) Spatially resolved Ar+ ion distribution produced by the 1-XUV-photon ionization process at the focus of the XUV beam. The XUV beam contains the 11[th] to15th harmonics with approximately equal amplitudes and a small amount of the 17th harmonic. (b) XUV beam profile in which the contribution of the long- (L) and short- (S) trajectory harmonics is visible due to their different divergence. $\varphi^L$ and $\varphi^S$ denote their phases. From ref. [64]. (c) Contour produced by the 1-XUV-photon ionization signal of Ar at different intensities $I_L \approx 2$- $7\times10^{13}$ W/cm². Upper-panel: The plot shows how the structure of the ion distribution along the z-axis of (a) is changing from single- to double- peak. The plot was produced after normalization of the line outs of the ion distribution along the z-axis at y = 0. The black line depicts the mean value of the ion distribution. The error bars represent the standard deviation of the mean value resulting by taking into account the accuracy of the laser intensity measurement. Lower-panel: Calculated (and normalized) contour plot showing the dependence of the z-axis interference pattern on $I_L$. (d) Measured phase differences between the short- and long-trajectory harmonics (upper-panel) from which the emission time differences have been obtained (lower-panel).The vertical black dotted lines depict the harmonic cut-off region. From. ref. [43].

The change of the ion distribution from single- to double-peak structure with $I_L$ is clearly shown in the measured contour plot of Fig. 6c, which is found to be in fair agreement with the theoretical calculations (Fig. 6d). The difference between the harmonic emission times ($\Delta t_e = t_e^L - t_e^S$) can be obtained from the measured contour plot by using the following considerations: (a) $\Delta\varphi^{S,L} = 0$ at lower values of $I_L$, where an $I_L$ independent single-peak structure is observed. This is in agreement with the harmonic generation theory, where in the cutoff region the two trajectories degenerate into one with a single phase. (b) $\Delta\varphi^{S,L}$ is increasing monotonically with $I_L$. (c) The $\Delta\varphi^{S,L}$ is increasing by $\pi$ when the structure changes from a single to a double peak. Following these considerations, the values of $\Delta\varphi^{S,L}$ with differences $n\pi$ ($n = 0,1,2,\ldots$) and $\Delta t_e = nT_q/2$ (where Tq is the carrier frequency of the plateau harmonic superposition) as a function of $I_L$ have been obtained and are shown in Figs. 6d (details can be found in ref. [43]).

**4.2. FEL XUV sources**

Similar results and information is obtained using FEL XUV sources for atomic ionization. Here due to the high XUV pulse energies very high charge states can be reached, the spatial distribution of which can be recorded with the IM. Such distributions obtained at FLASH at 13.7 nm XUV wavelength, µJ level pulse energies after focusing with a 60cm focal length multilayer mirror are shown in figure 7. The z-x distribution (z being the beam propagation axis) of Xenon charge states 2+ to 7+ generated in a 300 µm long portions of the XUV focus around the beam waist are shown in figure 7a. The radial extent of the spatial ion distributions is in the order of few tens of µm and decreases with increasing charge state. In figure 7b the width of the measured $Xe^{4+}$ ion distribution is plotted as a function of z. Since $Xe^{4+}$ is produced in a two-photon ionization process, assuming no distortion of the ion distribution due to space charge effects or depletion, the radius of the $Xe^{4+}$ distribution is a factor $\sqrt{2}$ smaller than that of the intensity distribution of the XUV beam. The blue circles are measured data, while the black curve is a fit of the beam size function $w(z) = w_0\sqrt{1 + z^2/z_R^2}$ to the measured data, with the beam waist $w_0 = w(0)$ and the Rayleigh length $z_R$ as fit parameters. This proof-of-principle study was the first demonstration of the applicability of ion microscopy to in-situ diagnostics of focus FEL beams.

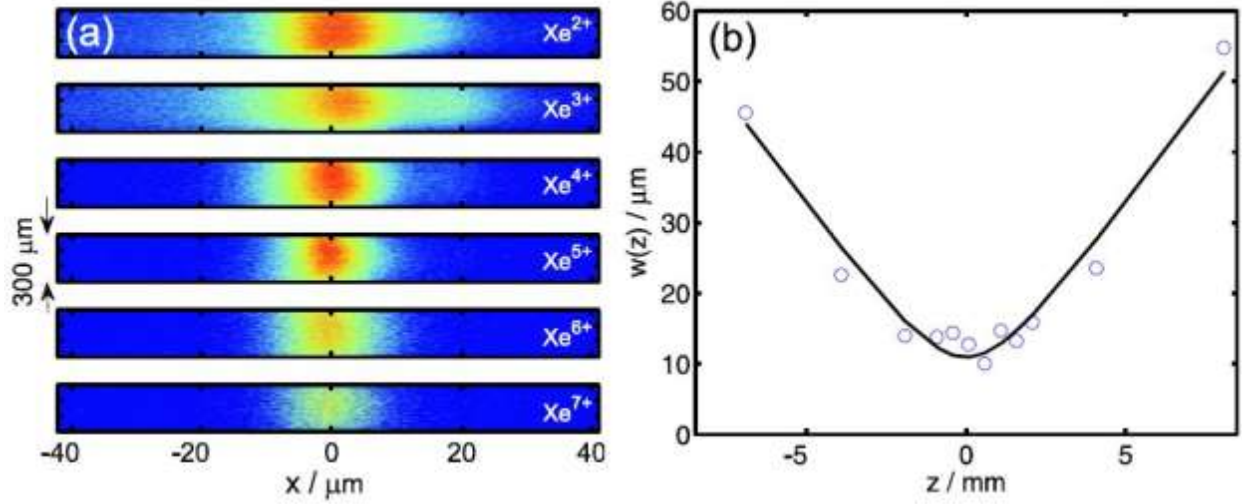

**Figure 7.** (a) Spatial distribution of the charge states $Xe^{2+}$ to $Xe^{7+}$ recorded at the beam waist. The beam propagation axis is vertical. (b) Blue circles: plot of the width w(z) of the $Xe^{4+}$ ion distribution (FWHM) as a function of the position z along the beam propagation axis. Black line: fit of the function $w_0\sqrt{1 + z^2/z_R^2}$ to the experimental data, with $w_0 = w(0)$ and the Rayleigh length $z_R$ as fit parameters. From ref. [41].

## 5. Single-shot time resolved studies in XUV spectral range with few-fs resolution

XUV-pump-XUV-probe schemes, as fully perturbative two-photon processes ($\gamma > 1$ for XUV intensities up to $\approx 5\times 10^{16}$ W/cm$^2$), have notable advantages in studies of ultrafast dynamics as compared to the IR-XUV pump-probe schemes. This is because the XUV pulse does not effectively distort the system (and its dynamics) under investigation, which is the case in some IR-XUV pump-probe schemes, as the number of photons inducing an XUV-pump-XUV-probe process are much less compared to those of an IR-XUV-pump-probe scheme.

However, an XUV-pump-XUV-probe measurement suffers from the intrinsic limitations that accompany any pump-probe approach that involves an interferometer (or an XUV-wave front beam splitter [34,35]) as a delay line between the pump and the probe pulses. In a pump-probe experiment the ultrafast evolution of the system is obtained by multiple shot measurements at different time delays between the pump and the probe pulses. During these measurements all the XUV-IR-laser parameters (XUV-pulse duration, XUV-intensity, Carrier-Envelope-Phase (CEP) of the IR pulse e.t.c.) and experimental conditions must remain stable. Additionally, a pump-probe measurement with few-*fs/asec* temporal resolution suffers from limitations in spectroscopic resolution due to complications on maintaining the *asec* resolution and the experimental parameters constant for long data acquisition times and long delay values between the pump-probe pulses. The generation of high-power few-*fs/asec* pulses requires the use of high-power *fs* laser systems operating at repetition rates from tens-of-Hz (conventional laser systems) up to 1kHz (state of the art systems) [65]. Focusing these pulses into a gas phase medium, XUV intensities $I_{XUV}$ higher than $10^{13}$ W/cm$^2$ can be achieved and observable non-liner processes can be induced [45]. In a XUV-pump-XUV-probe experiment, in order to obtain a spectral resolution ($\delta v$) of about 300 GHz (i.e. energy resolution ≈ 1 meV) and temporal resolution of about 100 asec (i.e. delay step between the pump and the probe pulses ≈7.5 nm) the total displacement between the pump and probe beams should be about 450 μm. This corresponds to a delay value of ≈ 3 picosecond (*psec*). Assuming a repetition rate of 10 Hz, and accumulating 100 shots for each delay value, the total data acquisition time would be about 90 hours. A tremendously large experimental effort is required for maintaining constant experimental conditions during such acquisition times. Besides this, there are additional intrinsic complications which do not allow to perform pump-probe studies in the few-*fs* time scale with GHz spectral resolution. They originate from the spatial displacement between the pump and the probe XUV foci positions at the interaction point [66]. This displacement results in a dramatic change of the intensities between the pump and the probe pulses during the measurement, making the analysis and the interpretation of the experimental results extremely complicated if not impossible.

We can overcome all these obstacles by utilizing an approach which provides high temporal and spectral resolution in a single-shot measurement. This is feasible by imaging the ion distribution, produced by a two-photon process, along the propagation axis of two focused counter propagating intense XUV pulses. The principle of the approach has been reported 20 years ago [67,68] in studies tracing the dynamics of "slow" wave packets of coherently excited high lying Rydberg states. The extension of the approach to time scales of few-*fs* or less was not feasible due to the lack of (i) intense ultrashort XUV pulses and (ii) high spatial resolution time-of-flight ion imaging detectors. High power XUV pulses with duration ≤ 2 *fs*, capable to induce two-photon processes, have been already produced [32,45,69] and implemented to XUV-pump-XUV-probe studies of 1-*fs*-scale wave packet dynamics in atoms [32,45] and molecules [70]. In addition, as is shown in section 3.1, images of the spatial ion distribution produced in the XUV focal area via linear and non-linear processes in atoms have been recorded by means of an ≈ 1 μm spatial resolution IM.

A schematic of a high spatial resolution single-shot two-photon correlator is shown in Fig. 8a.

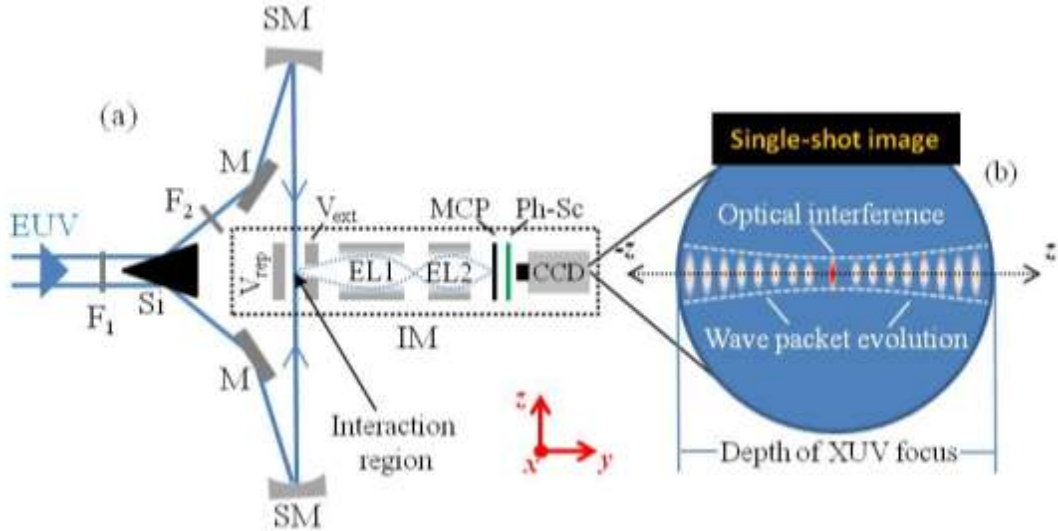

**Figure 8**. a) A schematic of the high spatial resolution single-shot two-XUV-photon correlator, showing its XUV optical set-up and the IM. Si, F1,2, M and SM are a Silicon wedge, thin metal filters, mirrors and spherical mirrors, respectively. b) A sketch of the imaged ion distribution along the propagation axis of the counter propagating pulses showing the areas where the optical interference ($z \approx 0$) and the wave packet evolution ($z < 0$ and $z > 0$) is taking place.

A p-polarized high power XUV beam propagates towards the single-shot two-XUV-photon correlator. A Si wedge manufactured with its surfaces at grazing incidence angle is placed at the entrance of the correlator in order to split the XUV beam and substantially attenuate the IR pulse. The reflected by the Si wedge XUV beams pass through the thin metal filters (F1, 2) which are used for the XUV spectral selection. Then, the two XUV beams propagate towards the arms of the correlator using broadband high reflectivity XUV mirrors (M). Subsequently, the XUV beams are focused in the interaction region of the IM by the spherical mirrors (SM). A sketch of the ion distribution produced by a two-XUV-photon ionization process in the region where the counter propagating XUV pulses interact with the medium is shown in Fig. 8b. It is assumed that the optical interference is taking place at $z \approx 0$ and the wave packet evolution appears at $z < 0$ and $z > 0$. The applicability of the approach has been tested by recording the $Xe^+$ ion distribution resulting from multi-photon ionization of xenon atoms in the field of two counter propagating few-cycle mid-IR laser pulses with a central wavelength of 2.1 μm (Fig. 9).

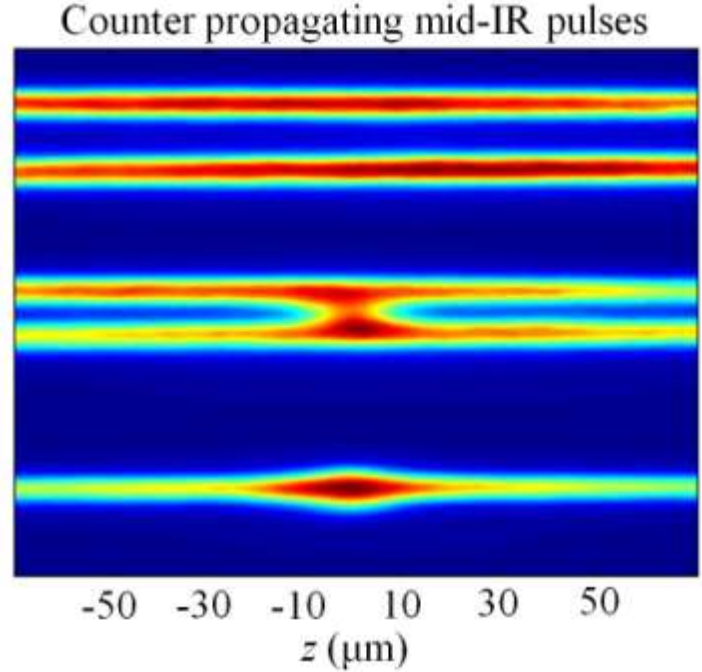

**Figure 9.** $Xe^+$ ion distribution generated in the focii of two counter propagating mid-IR pulses. Upper panel: when the focii are not spatially overlapped. Middle panel: when the two focii are partially overlapped at the edge of the beam cross section. Lower-panel: when the focii are spatially overlapped. The images of each panel are normalized to the maximum value corresponding to the dark red-color. In all panels the temporal overlap is at $z \approx 0$.

Moreover, the aforementioned approach has already been successfully implemented in a single-shot VUV-pump / VUV-probe scheme for the investigation of ultrafast molecular dynamics [71, 72]. Few-fs dissociation dynamics of $O_2$ and $H_2O/D_2O$ have been investigated within the respective ionization windows at 162 nm, utilizing an intense, sub-20 fs, fifth-harmonic (162 nm / 7.65 eV) pulse [73]. A single-shot intensity autocorrelation measurement of the fifth-harmonic pulse, obtained with the exact experimental setup, has enabled accurate deconvolution of the respective time-constants, disclosing dynamics beyond the instruments temporal resolution in the sub-10 fs range in both cases.

Using the specifications of the IM as have been described in section 2, a ≈ 500 μm long ion distribution can be resolved with 270 nm resolution i.e the wave packet evolution can be resolved with 2 *fs* resolution over a few-psec delay range resulting to a spectral resolution of ≈ 1 meV. The feasibility of recording a single-shot image induced by a two-XUV-photon process (hereafter will be called "non-linear single-shot image"), using few-fs XUV pulses with power of about 1 GWatt/pulse ($\tau_{XUV} = 1$ fs and $E_{XUV} = 1$ μJ/pulse), can be evaluated taking into account the conditions at which the images of section 3.1 have been recorded. These images have been measured using an XUV beam with power ~0.01 GWatt/pulse. In this experiment the XUV beam was focused with an f = 5 cm spherical mirror into the interaction area which was filled with Argon or Helium. The peak intensity of the XUV at the focus was ~ $10^{14}$ W/cm$^2$. For Helium, a "non-linear image" has been recorded accumulating ≈ 15000 shots. This image can be used as a reference is estimating the quality of the images which will be obtained using the ≈ 1 GWatt/pulse XUV pulse. By focusing a split XUV beam with peak power ≈ 0.5 GWatt into the interaction region with a spherical mirror of $f = 40$ cm, intensities up to $I_{XUV} \approx 10^{15}$ W/cm$^2$ can be reached resulting to an "non-linear single-shot image" with signal ($Y \propto I_{XUV}^2 \tau_{XUV}$) ~ 10 times stronger compared to the "non-linear image" of Fig. 8b. Such a signal level is sufficient for performing single-shot XUV-pump-XUV-probe studies. The number of ions that can be generated by a two-XUV-photon process within a volume of V ≈ $270^3$ nm$^3$ (this value corresponds to the volume which is resulted by an IM with ≈ 270 nm resolution) can be roughly estimated by the relation $Y = \sigma^{(2)} \cdot (I_{XUV}/\hbar\omega_{XUV})^2 \cdot \tau_{XUV} \cdot V \cdot \rho$ (where $Y$ is the ion yield, $\sigma^{(2)} \approx 10^{-48}$ cm$^4$s and $\rho$, are the non-resonant two-XUV-photon cross section and the gas density in the interaction region). For typical (in case of using pulsed gas jets) gas densities $\rho \approx 10^{15}$ particles/cm$^3$ and XUV pulses with photon energy $\hbar\omega_{XUV} \approx 20$ eV, duration $\tau_{XUV} \approx 1$ fs, and $I_{XUV} \approx 10^{15}$ W/cm$^2$, the number of ions generated within the $270^3$ nm$^3$ volume is ≈ 2x10$^3$ per pulse. This value is sufficient for the realization of this type studies. Also, the large amount of the generated ions provides sufficient flexibility for finding the optimum spatial resolution of the IM by reducing the energy of the XUV pulse or the gas density in the interaction region.

The temporal evolution of the wave packet can be directly obtained from the measured single-shot image induced by the counter propagating XUV beams. Each position of the image along the propagation axis (*z*) corresponds to a specific delay value between the pump and the probe pulses according to the relation $\Delta T = 2z/c$ (where $\Delta T$ is the difference in the pulse delay for particles which are separated by a distance *z*). The ion-distribution (Fig. 8b) at *z*=0 (zero delay between the pump and the probe pulses) contains the information about the optical interference (i.e. duration of the XUV pulses when that the pump and the probe pulses are identical) between the pump-probe pulses. For example, for an IM with spatial resolution of ≈ 270 nm, XUV pulses with duration down to ≈ 2 *fs* can be measured. The modulation of the ion distribution (Fig. 8b) at z < 0 and z > 0 reflects the wave packet evolution. In case of using identical XUV pulses the image will be symmetric with respect to $z = 0$ position, while in case of using different XUV pump-probe pulses the image will be asymmetric. Here, it is important to note that in case of mechanically ultra-stable configuration the correlator can be used in a many-shot operation mode. Also, in order to avoid having aberrations in the focus of the XUV beams, great care should be taken to keep the angle of incidence on the spherical mirrors < 2 deg. Additionally, the choice of the SM focal length $f$ is of crucial importance since it is associated with the aberrations of the XUV at the focus, the intensity of the XUV source in the interaction region and the spectral resolution ($\delta v$) of the measurement which is inversely proportional to the maximum delay values at which the wave-packet evolution can be studied. Furthermore, it should be ensured that both spherical mirrors are manufactured with the exact radius of curvature (spherical irregularity < $\lambda/20$, with $\lambda$ = 800 nm) in order to avoid a mismatch of their focal properties and thus compromise the spatiotemporal overlap of the two counter propagating beams. Using a SM with $f \approx 40$ cm, a pulse power of ≈ 1 GWatt and an XUV beam diameter of ≈ 3 mm, the XUV intensity in the focal area will be $I_{XUV} \approx 10^{15}$ W/cm$^2$. This intensity can produce a sufficient amount of ions via two-XUV-photon ionization process along a field of view ≈ 500 μm. Assuming that the measurable two-XUV-photon ion signal is mainly produced within the depth of the focus of the XUV beams, this distance will allow a measurement of the wave packet evolution for times in the 3 *psec* range resulting to a spectral resolution of ≈ 1 meV.

Summarizing this section, the spatially resolved ion distribution is the core of a single-shot 2$^{nd}$-order correlator, for spectroscopic studies with spectral and temporal resolution in the 1 meV and 1 *fs* range, respectively.

## 6. Single-shot 2$^{nd}$-order autocorrelator for the temporal characterization of asec XUV pulses.

As mentioned above in order to use the counter propagating approach for attosecond pulse metrology the resolution of the IM has to be better than the geometrical width of the *asec* pulse (100 *asec* = 30 nm)). Thus, as the present technology can provide IM with spatial resolution of 270 nm, (which translates to a temporal resolution at best 1 fs), neither attosecond pulse trains can be visualized nor attosecond pulse durations can be measured. Sub-fs temporal resolution requires a different geometry, namely crossed XUV beams, in which the smaller the crossing angle becomes the higher the temporal resolution one can reach.

The principle is very similar to the principle of the single shot 2$^{nd}$ order IR autocorrelator illustrated in Fig. 10. Here two IR beams at crossing under an angle θ in a non-linear crystal (BBO in the figure), where second harmonic is generated and its spatial profile is observed through a CCD camera. The duration of the IR pulse is linearly mapped in the spatial extent of the CCD image. Once the device is calibrated, measuring the size of the image one can deduce the pulse duration in straightforward manner.

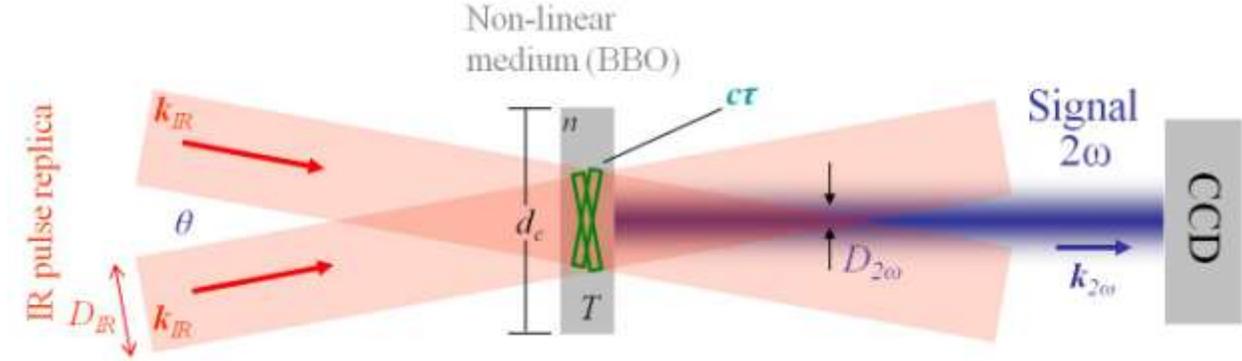

**Figure 10.** Operation principle of the conventional single-shot autocorrelator for IR fs laser pulses. $D_{IR}$ and $k_{IR}$ are the diameter and wave vector of the two IR laser pulses crossing each other in the BBO crystal, respectively. $n$ is the refractive index of the medium and $\tau$ is the duration of the IR pulses. $\theta$ is the angle between the two IR pulses. $d_c$, $T$ is the diameter and the thickness of the BBO crystal where the second harmonic (2ω) is generated. $D_{2\omega}$, $k_{2\omega}$ is the width and wave vector of the generated 2ω beam detected by the CCD camera.

Important requirements for a single-shot IR autocorrelator are: 1) The non-linear medium should be essentially dispersionless, introducing low GVD and having a spectrally flat non − linear response, in order to avoid pulse duration measurement distortion; 2) The diameter $D_{IR}$ of the IR beams on the crystal should be where $\tau$ is the duration of the pulse and $c$ the speed of light. This condition is obtained when the delay between the two pulses is zero at the crossing point.; 3) The thickness $T$ of the crystal should be $T \geq \tau \cdot c/[cos(sin(\theta/2)/n)]$, n being the refractive index of the crystal; 4) The diameter of the crystal should be $d_c \geq [D_{IR}/cos(\theta/2)]+T\,tan(sin(\theta/2)/n)$; 5) The spatial resolution of the 2ω detector (CCD) should be $CCD_{pixel\text{-}size} < D_{2\omega}/3$ (where $D_{2\omega}=c\cdot\tau/[sin(\theta/2)]$ is the width of the 2ω beam on the CCD) in order the spatial distribution of the 2$^{nd}$ harmonic to be recordable. Conditions 2, 3 and 4 imply that the dimensions of overlapping pulses are smaller than the corresponding crystal dimensions and thus the spatial profile of the generated 2$^{nd}$ harmonic results from the entire temporal distribution of the pulse and not from an artificially truncated one.

This technique can be applied in the XUV spectral region if the second harmonic process is replaced by a two-XUV-photon ionization process and the CCD camera by an ion microscope recording the spatial ion distribution produced by the two overlapping XUV beams via non-resonant two-photon ionization. The method is very similar to the 2$^{nd}$ order single shot AC of *fs* pulses. There is though the following important difference between the two methods. The single shot AC method shown in Fig. 10 is background free as only the 2nd harmonic produced along the bisector of the two crossed IR beams is detected while in the two-XUV-photon ionization process, ions are produced and detected wherever the XUV intensity is sufficiently high, even if only one beam is present and thus the measured image includes background. The principle of the method is depicted in Fig. 11.

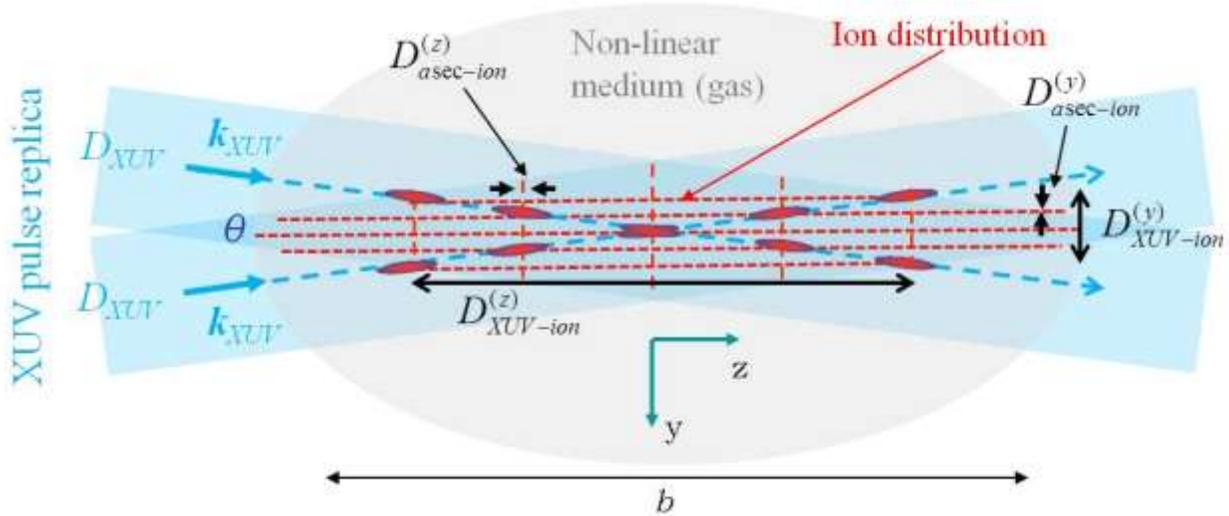

**Figure 11.** Operation principle of the single-shot asec XUV autocorrelator. $D_{XUV}$, $k_{XUV}$ is the diameter and wave vector of the two XUV laser pulses. The grey shaded area depicts the gas medium. $\theta$ is the angle between the two XUV beams. In this sketch we assume that the 2-XUV-photon ion distribution is produced only in a length equal to the confocal parameter of the XUV beam (*b*) where, for reasons of simplicity, the intensity is assumed to be constant. $D^{(y)}_{XUV-ion}$, is the overall width of the ion distribution (produced by a 2-XUV-photon ionization process) along the y-axis which is directly related to the overall duration of the XUV pulse. The red dashed lines are the ion distributions produced by an asec pulse train. The width of these lines $D^{(y)}_{asec-ion}$ is directly related to the duration of the asec pulses in the train. Similar is the case regarding the ion distribution along one of the two beams (z'-direction). $D^{(z')}_{XUV-ion}$ is the total width of the ion distribution along the z'-axis and $D^{(z')}_{asec-ion}$ is the width of the ion distribution produced by the *asec* pulses in the train. Although the measurement of the ion distribution in one of the two directions is sufficient for the characterization of the asec pulses (as in the case of the conventional fs pulse metrology approach shown in Fig. 9), the two dimensional (y-axis and z'-axis) ion distribution (red-filled ellipse) can provide information about the evolution of the asec pulses in the medium.

In an interferometric 2$^{nd}$ order AC, along the y-axis the two anti-parallel $k^{y}_{XUV}$ components create "standing" waves, which result into an ion distribution consisting of lines parallel to the bisector of the two beams. The modulation period along the y-axis is $\lambda_{IR}/4\sin(\theta/2)$. This is a rather small period produced by the carrier frequency of the XUV pulses that cannot be resolved by the ion microscope, i.e. an interferometric AC is not recordable. For attosecond pulse trains (ATP) an additional modulation of the ion signal comes for the $\lambda/2$ spacing between two subsequent attosecond pulses. For an APT the spacing between the maxima of the ion distribution along the y-axis produced by the envelope of the attosecond pulses (intensity AC) is $d = \lambda_{IR}/4\sin(\theta/2)$ and depending on the geometry that is recordable. The duration of the attosecond pulses can be deduced from the width of these maxima to be, $\tau = D^{(y)}_{asec-ion}\sin(\theta/2)/c$, $D^{(y)}_{asec-ion}$ being the width of the ion distribution along the y-axis produced by the *asec* pulses in the train. Thus, in order some pulses of an APT to be imprinted in the measured image $d < w_0^{XUV}$, $w_0^{XUV}$ being the XUV beam waist. For the high XUV intensities required for the observation of a 2-XUV-photon process, beam waists of 10 μm or less i.e. small crossing angles $\theta \leq 2^o$ are required. In order isolated attosecond pulses to be measurable the spatial extent of the ion distribution they produce has to fit in the beam waist. The requirement $D^{(y)}_{asec-ion} \approx w_0^{XUV}$ implies that for $w_0^{XUV} = 10$ μm, a crossing angle of $2^o$ and a spatial resolution of the ion microscope of 1 μm the attosecond pulse durations $\tau$ that can be measured are $60 \leq \tau \leq 600$ asec. An increase in the crossing angle by few degrees will increase both limits almost linearly.

In case there is enough signal for recording the y-z plane image of the entire focal area a modulation of the ion distribution along each of the two beams can measured. The modulation period is $\lambda_{IR}/4\sin^2(\theta/2)$. While this distribution does not add additional information about the duration of the pulse it is significantly magnified with respect to the modulation along the y-axis. The pulse duration can then be deduced from this distribution to be $\tau = D^{(z')}_{asec-ion}\sin^2(\theta/2)/c$, $D^{(z')}_{asec-ion}$ being the width of the ion distribution along one of the beams, produced by the *asec* pulses in the train.

An additional requirement for the non-linear autocorrelator is the ionization process to be "instantaneous" and spectrally flat. This implies that the 2-XUV-photon ionization process should be non-resonant with states having lifetimes comparable to the duration of the pulse to be characterized. Requirements that must be fulfilled for a single shot autocorrelation are: $D_{XUV} \gg D^{(y)}_{XUV-ion}$ (an acceptable value is $D_{XUV} > 2\, D^{(y)}_{XUV-ion}$), $D_{XUV}$ being the diameter of the XUV beams, $b \gg D^{(z')}_{XUV-ion}$ (an acceptable value is $b > 2\, D^{(z')}_{XUV-ion}$), $b$ being the confocal parameter of the XUV beam and the spatial resolution of the ion microscope must be better than $D^{(y,z\prime)}_{asec-ion}/3$.

A detailed description of the single shot non-linear XUV autocorrelator principle can be found in [74] where crossed beams geometries have been assessed. Those are introducing a two pinhole mask or a two slit mask in the XUV beam or splitting the XUV beam into two and focusing the two parts in a common focus. The later geometry has been found to be the most efficient and is summarized here. The splitting occurs using a wedge as shown in Fig. 12. In this figure the wedge material is Si but it can be any other material that reflects the XUV and absorbs or transmits the IR. Thus the wedge may act as a beam divider and XUV-IR separator at the same time. It is worth noting that for small crossing angles (few degrees) the angle of incidence substantially deviates from the $75^o$ Brewster angle of Si for 800 nm and thus the IR suppression will be small. However, for larger crossing angles allowing measurement of durations of few hundreds of attoseconds the IR suppression can be notable (1-2 orders of magnitude). Depending on the XUV photon energy the geometry can be as shown in Fig. 12. For low photon energies near normal incidence reflection of the two XUV beams by a gold coated or multilayer spherical mirror may be used in a rather simple geometry. For higher photon energy grazing incidence geometries have to be used. An example is outlined in Fig. 12b, where the grazing incidence reflecting optics can be toroidal or ellipsoidal mirrors.

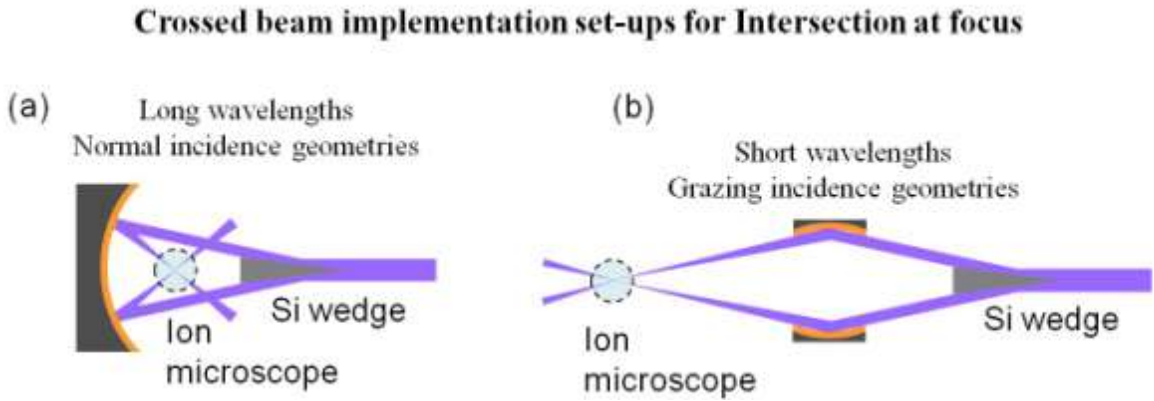

**Figure 12.** Geometries producing two XUV beams intersecting at their focus. The arrangement in (a) is appropriate for low photon energies < 100 eV for which near normal incidence reflection from gold coated or multilayer spherical mirrors is efficient. For larger photon energies grazing incidence optical arrangements have to be used. In (b) the mirrors can be toroidal or ellipsoidal ones.

Experimentally important steps have been made towards a XUV single shot $2^{nd}$ order non-linear AC. The feasibility of observing spatially resolved two-XUV-photon ionization of He has been established in a recent experiment [44]. In another experiment the spatial distribution of ions produced by one-XUV-photon ionization of Ar using two crossed XUV beams has been recorded. The experiment has been performed using the two pinhole mask introduced in the XUV beam discussed above. The arrangement is shown in Fig. 13. The XUV beam of a comb of harmonics 11-15 (train of pulses) produced by a ≈ 33 *fs*, ≈ 15 mJ Ti:Sapphire beam focused with a 3m long focal length spherical mirror in the Xe jet of a pulsed nozzle is impinging on a mask with two pinholes of ≈ 120 μm diameter in ≈ 3 mm distance from each other. The two transmitted XUV beams are reflected and focused by a gold coated spherical mirror in a Ar jet. They intersect each other at the focus under an angle of ≈ 3.5$^o$ ionizing Ar. For the harmonics used ionization of Ar is a single photon process. Results are shown in the lower panels of Fig. 13. In the left panel the two crossed beams are visualized by the image ion distribution.

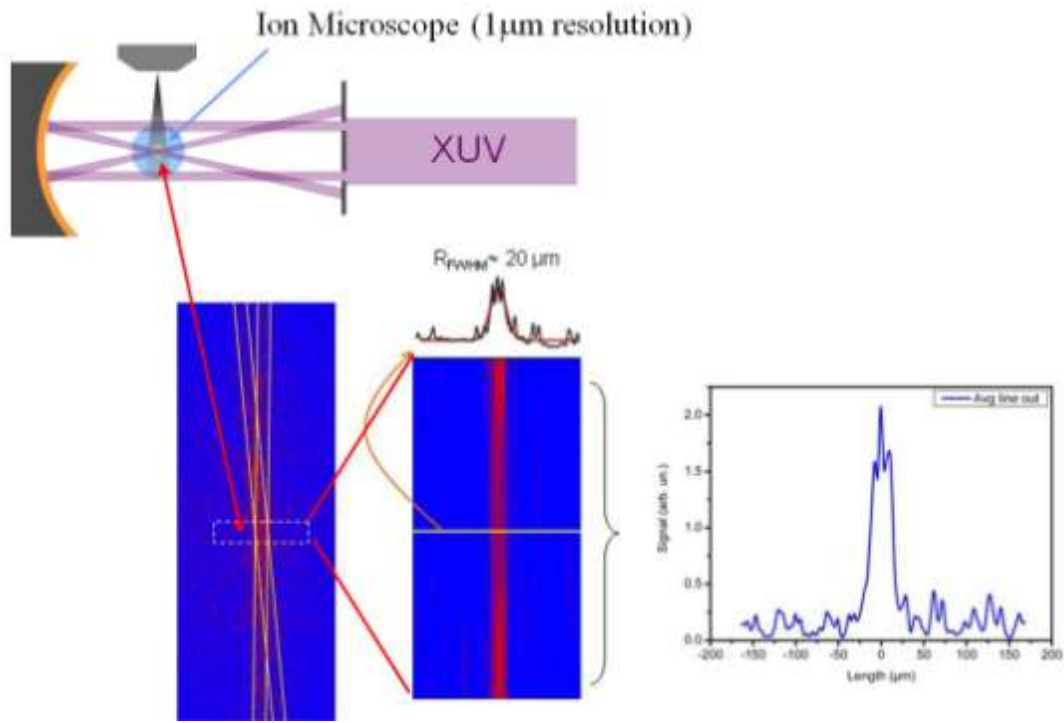

**Figure 13.** Single photon ionization of Ar by two crossed XUV beams (pulse trains). In the upper panel is shown the experimental set up. A mask with two pinholes of 120 μm diameter is inserted in the XUV beam. The two formed beams are impinging on a gold coated spherical mirror. The reflected beams are focused and they cross each other at the focus. The measured image of the Ar ion distribution is shown in the lower-left panel. The lower-middle panel shows the ion distribution of the yellow-squared-dashed area of the lower-left panel. The curve on the top of the middle panel is the line-out signal across the y axis (yellow line perpendicularly to the propagation axis z). The curve at the lower-right-panel is the integrated signal of the lower-middle panel..

The plane shown is the y-z plane. The panel in the center is an expanded view of the focus. The FWHM radius of the focus is measured to be ≈ 20 μm. The right panel is the integrated signal of the middle panel. Although the distribution is resulted by an one-XUV-photon ionization process, an indication of modulation in the ion distribution is present. However, the modulation is not clear enough for a quantitative evaluation of this periodic structure. For the crossing angle used the resolution is not enough for a quantitative analysis. The above two experiments are first steps towards a pulse temporal characterization through the described method. Implementation of the geometry of Fig. 12a in a two-XUV-photon ionization process is the next step towards the single shot $2^{nd}$ order AC measurement goal.

## 7. Conclusions

We have demonstrated how spatially resolved photoionization yields resulting from light-matter interactions can be used to obtain detail information about I) the properties of the light source, II) the ultrafast dynamics of the ionization process in atoms/molecules. We have reviewed the operation principle of the ion microscopy approach, which has been recently developed to obtain images of ion distributions with high spatial resolution, potentially down to the sub-μm scale. The device has found applications in the quantitative study of linear and non-linear light matter interactions in focused IR and XUV beams, in single-shot as well as in multi-shot operation. Further, we have discussed the applicability of the approach in performing single-shot time autocorrelation measurements of *asec* pulses. It is worth noticing that the IM technology could also be combined with the streaking method, allowing detection of ionization in very small time intervals. This combined technology would require a device that can image the electron distribution at the focus of the beam and would thus provide simultaneous temporal, spatial and energy resolution.

**Disclosure statement**

No potential conflict of interest was reported by the authors.

**Funding**

This work has received funding from the Greek funding program NSFR and the European Union's Horizon 2020 research and innovation program under Marie Sklodowska-Curie grant agreement no. 641789 MEDEA and Laserlab-Europe, H2020 EC-GA 654148. ELI-ALPS is supported by the European Union and co-financed by the European Regional Development Fund (GINOP-2.3.6-15-2015-00001).